\documentclass{aa}  
\usepackage{graphicx}
\usepackage{natbib}
\usepackage{amssymb}
\usepackage{amsmath}
\usepackage{amstext}
\usepackage{rotate}
\usepackage{rotating}
\usepackage{supertabular}
\usepackage{epsfig}
\usepackage{lscape}
\usepackage{enumerate}

\bibpunct{(}{)}{;}{a}{}{,}
\usepackage{txfonts}
\newcommand{\gsim}{\hbox{\rlap{\lower.55ex\hbox{$\sim$}} \kern-.3em
\raise.4ex \hbox{$>$}}}
\newcommand{\lsim}{\hbox{\rlap{\lower.55ex\hbox{$\sim$}} \kern-.3em
\raise.4ex \hbox{$<$}}}

\begin{document}
%
   \title{Search for Tidal Dwarf Galaxies Candidates
  in \\a Sample of Ultraluminous Infrared Galaxies}
   \author{A. Monreal-Ibero\inst{1}
          \and
          L. Colina\inst{2}
          \and
          S. Arribas\inst{2}
          \and
          M. Garc\'{\i}a-Mar\'{\i}n\inst{2}
          }

   \offprints{A. Monreal-Ibero}

   \institute{Astrophysikalisches Institut Potsdam, An der Sternwarte 16
              D-14482 Potsdam\\
              \email{amonreal@aip.de}
         \and
             Instituto de Estructura de la Materia (CSIC), Serrano 121, 28006
              Madrid\\ 
             \email{[colina,arribas,maca]@damir.iem.csic.es}
             }

   \date{version ami 25/05/2007}

 
  \abstract
   {Star-forming small galaxies made out of collisional debris
  have been found in a variety of merging systems. So far only a few of
  them are known in Ultraluminous Infrared Galaxies (ULIRGs) although
  they show clear signs of interactions. Whether external star
  formation may take place in such objects in an open question.}
   {The aim of this paper is to identify and characterise the
  physical and kinematic properties of the external star
  forming regions  in a sample of ULIRGs, including TDG
  candidates. The likelihood of survival of these regions 
  as TDGs is also evaluated.} 
   {The analysis is based on optical Integral Field Spectroscopy (IFS) and
   high angular resolution HST imaging.} 
   {We have found that the presence of external star-forming
   regions is common with 12 objects being identified in 5
   ULIRGs. These regions show a large range of dynamical mass  
   up to $1\times10^{10}$~M$_\odot$, with  average
   sizes of $\sim$750 pc. In addition, the line ratios (\textsc {H\,ii}
   region-like), metallicities ($12 + \log 
   \mathrm{(O/H)} \sim$8.6) and H$\alpha$ equivalent widths ($34 -
   257$~\AA)  are typical of young bursts of star formation  (age~$ \sim 5 -
   8$~Myr), and similar to those of the TDG candidates found in less
   luminous mergers and compact groups of galaxies. The extinction  corrected
   H$\alpha$ luminosity of these young bursts leads to masses for the
   young stellar component of $ \sim 2 \times 10^6 - 7 \times 
   10^8$~M$_\odot$.

The likelihood of survival of these regions as TDGs is
discussed based on their 
structural and kinematic properties. Particularly interesting
is our finding that  most of these  systems
follow the relation between  effective radius and velocity  dispersion found
at lower (globular clusters) and higher (Elliptical)  mass
systems, which suggests they are stable against internal
motions.
The stability against forces from the parent galaxy is
studied on the bases of several criteria and a comparison of
the data with the predictions of dynamical evolutionary models is
also performed. Five regions out of twelve show
\emph{High-Medium} or \emph{High} likelihood of survival
based on all the utilised tracers. 
Our best candidate, which satisfy all the utilized criteria,
is located in the advanced merger 
IRAS~15250+3609  and presents a velocity field decoupled from the
relatively distant parent galaxy.
}
 {}


   \titlerunning{TDG candidates in ULIRGs}
   \maketitle
%

\section{Introduction}

Ultraluminous Infrared Galaxies (ULIRGs) are defined as
  objects with an 
infrared luminosity $L (8-1000\mu \mathrm{m}) \ga 10^{12}
L_{\sun}$  (e.g. Sanders and Mirabel, 1996). This huge luminosity 
is believed to be caused mainly by star formation, although the
existence of an AGN cannot be discarded and could be the dominant
source of energy in a small percentage of these systems \citep[for
  reviews see][]{san96,lon06}.
ULIRGs are systems rich in gas and dust \citep[e.g.][]{eva02,eva00}
and all of them 
present emission lines in the optical
\citep[e.g.][]{vei99,kim98a}. Observational
 studies about the morphology of these systems indicate that most (if
 not all) of them show signs of mergers and interaction 
\citep[e.g.][]{cle96,bor00,bus02,cui01,vei02} while
theoretical simulations of mergers indicate that ULIRGs
constitute a well defined subgroup within mergers in general. That is, in  
principle only major mergers of two gas-rich spiral galaxies are
able to explain the 
large luminosity observed in ULIRGs and only during a short period of time
\citep[$\sim 50$~Myr, e.g.][]{mih96,bek01}.
In this context, there are some empirical evidences suggesting that
  ULIRGs are mergers of two or more (sub)-L$^\ast$ spirals \citep{col01}, that 
  could be the progenitors of intermediate-mass ellipticals
  \citep{gen01,tac02}. ULIRGs could also evolve into quasars  \citep{san88},
  at least in those mergers where the progenitor 
  galaxies are more massive \citep{col01}.

One issue regarding ULIRGs that deserves attention is the presence of
knots and condensations of star formation outside the nuclei. On the one
hand, at the lowest masses and smallest sizes range, we find the
so-called \emph{Super Stellar Clusters}
\citep[e.g.][]{sco00,shi01,sur00,sur99}. These 
systems, with typical 
masses in the range of $10^4 -10^6$~M$_\odot$ and sizes of $r \sim$5~pc, are
believed to be the progenitors of today globular clusters \citep[e.g.][]{sch96}
and, in addition to ULIRGs, they have been 
found in a wide range of environments such as starburst galaxies (see for
instance \citealt[][]{meu95} and \citealt[][]{mel05}), the space between
galaxies in compact groups \citep[][]{gal01,men04} and specially in mergers
less luminous than those studied here
\citep[see][and others]{zep99,whi99,alo02,kni03,bas06}. 
On the other hand, at masses $\sim 10^8 - 10^9$~M$_\odot$ 
we find the so-called \emph{Tidal Dwarf Galaxies}
(TDG). Their existence were already proposed in the 50's when
\citeauthor{zwi56} suggested that 
the tidal forces in interacting galaxies could
create self-gravitating objects made up from the debris of the interaction
that, perhaps, would become small galaxies. 
From the observational point of view, this 
kind of objects, with masses and sizes similar to dwarf galaxies and
large quantities of gas \citep[e.g.][]{bra01} have already been found
in less luminous interacting or merging galaxies 
\citep{duc94,duc98,duc00,hib01,wei00,wei03} or in compact groups
of galaxies \citep{igl01,men01,tem03,lop04,amr04,lin04} while modelling
has shown that the formation of condensations is possible in mergers between
two disc galaxies, as those happening in ULIRGs 
\citep[see][and others]{duc04,wet05,bou06}. 

In spite of the fact that they have the appropriate nature to
harbour TDG candidates, no systematic search for these 
candidates among ULIRGs has been performed up to date. The TDG
candidate in The Superantennae \citep{mir91}, turned out
to be a background object and to our knowledge, only three more
candidates have been identified so far \citep{mih98}.

An interesting question is
whether these objects are likely to contain the same quantity or more
TDGs than other types of merging systems. More generally, are ULIRGs
favourable sites for star formation in and extended mode?
Regarding the gas distribution, it seems that 
ULIRGs are particularly efficient in driving gas to the
innermost regions, meaning that there is a priori much less
gas reservoir in  the outer parts. However,
detailed simulations able to explain the luminosities
observed in these objects and focused in the gas distribution show a
dual behaviour for the gas: while the one in the inner disk 
(typically $\sim 5$~kpc) flows directly toward the central regions within
$1-2 \times 10^8$ yr after the pericentric passage, the outer gas is ejected
into tidal tails and bridges \citep{ion04}. Regarding the extend of
star formation in mergers, \citet{bar04} showed how this is larger
when shock-induced star formation laws are used instead of local gas 
density based laws.
From the observational point of view, the fact that
ULIRGs are more gas-rich than less luminous interacting systems
and the presence of shocks in these
systems \citep{mon06} may favor extended star-formation.
In this context, the positive or negative detection
of TDGs (or external star formation in general) may help to clarify
these issues.

Finally, quantifying the incidence of TDGs in local ULIRGs
may have cosmological implications since these objects are thought
to be the local 
counterpart of the so-called \emph{Spitzer} and sub-millimeter sources
\citep[e.g.][]{sma97,hug98} at $z \sim 1-2$. These sources present
similar or even greater luminosities than ULIRGs 
\citep[see ][ for a review of their properties]{bla02} and
a relatively high fraction of them have morphological
properties consistent with being systems suffering an interaction or
merging process (\citealt[][]{cha03,pop05,sma04,ivi00,con03}). They
are two orders of magnitude more numerous than local ULIRGs
\citep[][]{per05,cap07}. 
  Thus, while the present-day ULIRGs could provide a relatively small
  contribution to the total number of newly created TDGs, their high-$z$
  counterpart might be fundamental in this regard. Hence the detection of this
  kind of regions among ULIRGs may 
  increase considerably the estimation of the percentage of today dwarfs
  that could have been formed from tidal debris.

In this paper, we characterise the extranuclear star forming regions
of a sample of ULIRGs using the combined information of Integral Field
Spectroscopy (IFS) data together with high resolution images from the
HST. Properties such as extinction, ionisation state, metallicity, age and
mass of the stellar population, velocity dispersion, relative velocity etc.,
will be derived. We will use these parameters to
estimate the likeliness of survival of these regions as future TDGs. The
present work is part of a wider program whose final aim is performing
a detailed study of a representative sample of ULIRGs using IFS. Due
to the complex nature of these systems, this technique, that allows to
obtain at the same time spectral and spatial information, is well
suited for their characterisation.
Previous results of this program can be found in \citet{mon06} and references
therein.

Throughout the paper, a cosmology with  70~km~s$^{-1}$~Mpc$^{-1}$,
$\Omega_M =  0.3$ and $\Omega_\Lambda = 0.7$  is
assumed.


\section{Sample and Observations}

\subsection{Sample}

   \begin{table*}
\centering
      \caption[]{ULIRGs sample \label{misulirgs}}
              \begin{tabular}{ccccccccc}
            \hline
            \noalign{\smallskip}
Galaxy & z$^{\mathrm{a}}$ & Scale & $\log(L_{IR}/L_{\odot})^{\mathrm{b}}$ &
            IR Class$^{\mathrm{c}}$ & Interaction & Regions$^{\mathrm{e}}$\\   
       &  & (kpc arcsec$^{-1}$) &  & & class.$^{\mathrm{d}}$ \\
            \noalign{\smallskip}
            \hline
            \noalign{\smallskip}
IRAS~08572+3915   & 0.058 & 1.13 & 12.17 & W & \textsc{iii} & 2\\
IRAS~12112+0305   & 0.073 & 1.39 & 12.37 & C & \textsc{iii} & 5\\
IRAS~14348$-$1447 & 0.083 & 1.56 & 12.40 & C & \textsc{iii} & 1\\ 
IRAS~15250+3609   & 0.055 & 1.07 & 12.09 & C & \textsc{iv}  & 1\\ 
IRAS~16007+3743   & 0.185 & 3.10 & 12.11 & C & \textsc{iii} & 3\\
            \hline
         \end{tabular}
\begin{list}{}{}
\item[$^{\mathrm{a}}$] Redshifts taken from the NASA/IPAC Extragalactic
  Database (NED).
\item[$^{\mathrm{b}}$] Infrared luminosities calculated using the
  infrared IRAS fluxes of \citet{mos93}, the $L_{IR}$
  expression given in \citet{san96} and a luminosity distance assuming $H_0
  =70$~km~s$^{-1}$~Mpc$^{-1}$, $\Omega_M = 
  0.3$ and $\Omega_\Lambda = 0.7$.
\item[$^{\mathrm{c}}$] Standard IR classification where warm (W) ULIRGs have
  $f_{25}/f_{60} > 0.2$ while cold (C) ULIRGs have $f_{25}/f_{60} < 0.2$.
\item[$^{\mathrm{d}}$] Following the criterion proposed by
  \citet{vei02} where \textsc{iii} means pre-merger and \textsc{iv},
  merger state.
\item[$^{\mathrm{e}}$] Number of external star-forming regions
  analysed in this work. 

\end{list}
   \end{table*}
%


   \begin{figure*}
   \centering
\includegraphics[angle=0,scale=.93, clip=,bbllx=67, bblly=172, bburx=535,
bbury=788]{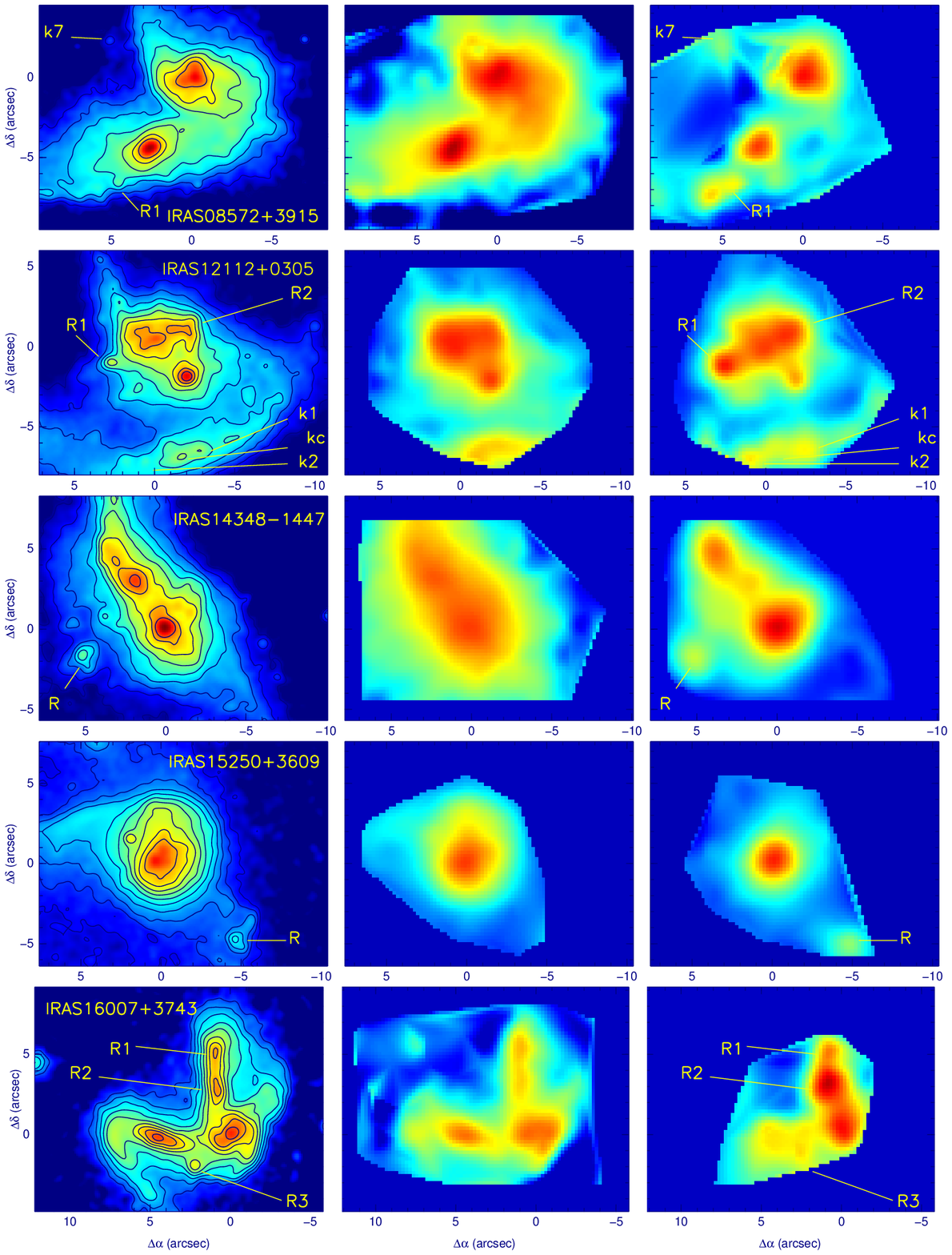}
   \caption[Sample of ULIRGs under study]{Sample of ULIRGs used in
  this study. First column displays the 
  WFPC2 image in the F814W filter (i.e. HST I-band) for
  reference. Images have been smoothed with a $0\farcs2$-sigma 
  Gaussian and contours have been over-plotted to better see the regions
  under study. Second and third columns present the maps
  created from the INTEGRAL data for a line-free continuum adjacent to
  H$\alpha$ and for this line respectively.
  All images are presented in logarithmic
  scale. North is up and east is to the left. Regions under study have been
  marked in both the WFPC2 and the INTEGRAL H$\alpha$ images to better see
  the matching between images from both instruments. \label{fig_sample}}
 \end{figure*}

The galaxies under study were selected from a sample of nine ULIRGs
with IFS data, including those presented in 
\citet{col05} plus IRAS~16007+3743 \citep{gar05}. They cover the ULIRGs low
luminosity range ($12.10 \le \log (L_{\mathrm{IR}}/L_\odot) \le 12.60$), have
a variety  of activity classes, and are in different stages of the interaction
process. 

Potential TDG candidates have been selected as
any high surface brightness 
compact region in the emission line maps (obtained from the IFS data) at a
distance from the nucleus of the galaxy larger than 2\farcs0, and 
associated with a single condensation in the WFPC2 image.  

Following the above selection criteria, no regions are
found in Mrk273 \citep{col99},
Arp220 \citep{arr01} and IRAS~17208$-$0014 \citep{arr03} in the IFS data. 
IRAS~15206+3342 presents a chain of knots at a projected distance of
about $3.5-7.5$~kpc from the nucleus in the WFPC2 image
\citep{arr02}. However, the 
spatial resolution of the IFS data doesn't allow us to derive their
individual  properties and, therefore, this system will not be considered here.

Thus, the final sample include the following five systems with at least one
region of interest: IRAS~08572+3915, IRAS~12112+0305, IRAS~14348$-$1447,
IRAS~15250+3609 and IRAS~16007+3743.
IRAS~15250+3609 is an advanced merger while the rest, with two nuclei and a
more spread star formation, 
are in an earlier stage of the merging process.
Table \ref{misulirgs} shows some relevant properties of the 
selected galaxies. With the exception of IRAS~16007+3743, which is at
a slightly higher redshift, typical redshifts for these systems are $z
\sim 0.07$ which implies a linear scale of about 1.3~kpc~arcsec$^{-1}$.

\subsection{Observations}

The IFS data  were obtained with the INTEGRAL system \citep{arr98} plus the 
WYFFOS spectrograph \citep{bin94} in the 4.2~m WHT at the Observatorio
del Roque de los Muchachos (Canary Islands) on April 1998 and April
2001. Spectra were taken using a 
600 lines mm$^{-1}$ grating with an effective resolution of 4.8~\AA.
We used the fibre bundle SB2 which has a field of view of  $16\farcs0 \times
12\farcs3$ and is  made up of 219 fibres, each 0\farcs45 in radius. Fibres are 
arranged in two sets which observe simultaneously the target and the
sky. The covered spectral range, exposure time and air mass for each object
can be found in \citet{mon06}. The data for IRAS~16007+3743 were
obtained with a similar configuration and with a total
exposure time of 6$\times$1500~s.  


\section{Data Reduction and analysis}

General reductions for the present IFS data have been discussed somewhere else
\citep[e.g][and references therein]{mon06}.  Here we will detail the
procedure for the absolute calibration of the 
data, which has not been included in previous papers in this series
and is of particular relevance for deriving the H$\alpha$ luminosity of the
regions under study.

\subsection{INTEGRAL data absolute flux calibration}

IFS seems to be a suitable technique to carry \textbf{out} the absolute flux
calibration in objects
like ULIRGs, with a complex morphological and kinematical  
structure and at a certain redshift. Other options have greater
difficulties. For example, observations with long-slit present
centring problems, worsened in presence of differential atmospheric
refraction. On the other hand, observations with narrow filters are
complicated to calibrate due to redshift, internal movements of the system
and blending of some lines, in particular H$\alpha$+\textsc{[N\,ii]}.

As a first step, the relative flux calibration was carried out using the fibre
with greater S/N in the calibration star image to create a sensibility
function which was 
used to calibrate \emph{every} spectrum, correcting from possible effects of
differential atmospheric refraction \citep{arr99,fil82}, and from the light
lost between fibres which affect the calibration star observations.  Note
that one can infer the fraction of the total flux collected by the
selected fibre from images generated from the IFS data at different
wavelengths (in a $\Delta\lambda$) . This information allows us to find
the conversion factors counts/absolute flux and, therefore, calibrate
absolutely the IFS data.     
In practise we proceed as follows. Firstly, we divided the calibration
star spectra in several 
sections of a few Amstrongs each. Secondly, using the flux measured 
for each fibre, and its position in the bundle, we created a flux map
of the calibration star for each section. Finally, we measured in that
map the flux collected by the selected fibre and the total flux of the
star in arbitrary units. The ratio between the flux collected
by the fibre to the total flux of 
the star in the map allows us to convert the counts from the star into absolute
flux in the selected wavelength. This is repeated for a set of  images at
different wavelengths, providing the conversion curve. In this specific case,
we obtained the conversion factor generating images (at
0.04~arcsec~pix$^{-1}$) in ten sections of 300~\AA~ each through the
entire observed spectral range. 

Three tests were performed in order to estimate the goodness of this
calibration method. Firstly, we compared the total measured flux in a certain
spectral range for an image of a star once calibrated with additional
calibration stars available for the night. Secondly, the internal
consistency of the calibration method was checked by comparing 
measurements in H$\alpha$ and H$\beta$ in certain 
emission line regions for one of our targets (IRAS~12112+0305) which had been
observed  in two different nights, pointings and rotation
angles. Finally, the
H$\alpha$ flux measured in the nuclei of this system were compared with
those found in the literature \citep{kim98a} simulating the aperture of an slit
in the interpolated maps.  From all these control tests we estimate 
uncertainties in the flux calibration  of about $10 - 15$\%.

\subsection{HST data}

Data archival images from the HST taken with the WFPC2 (F814W filter)
were also used to complete our analysis. All images but those of
IRAS~08572+3915  were taken in snapshot mode and
with the optional parameter \texttt{CR-SPLIT} activated. For every pointing
there were two already reduced images which were combined to reject cosmic
rays. Integrated counts for a given aperture were converted to 
magnitudes in the Vega system using the expressions given in \citet{bag02}.

F814W filter is similar to the Johnson-Cousin I filter. Transformations
between the WFPC2 and the Johnson-Cousin filter systems 
need at least measurements in two filters \citep[see for
example][]{ori00}. 
As in most of the regions, the only HST measurement available was that with
this filter, we have decided not to apply any transformation and to use
directly the F814W magnitudes as if they were I
magnitudes. We have performed a comparison between the expected
  magnitudes for the F814W and I filter with SYNPHOT using different  synthetic
  spectra. The more relevant for this study was the one corresponding to an
  instantaneous burst of 6~Myr with different levels of
  extinction. Differences were always 0.1 mag or less.




\section{Results \label{resultados}}

\subsection{Identification of the regions of interest}

The selected regions are identified as those bright condensations in the
emission lines outside the nuclear region (i.e. at distances of at
least 2~kpc from the nucleus). They are marked in the first
column of Figure \ref{fig_sample} which contains the WFPC2/HST images. The
other two columns in this figure show 
the continuum and H$\alpha$ emission line maps. 
The stellar continuum was obtained averaging two continuum ranges towards the
blue and red of the H$\alpha$ emission line simulating the action of a filter. 
For the ionised gas component, we have fitted Gaussian functions to
the emission lines using DIPSO package \citep{how88} within the STARLINK
environment \citep[details 
can be found in][]{mon06}. The general morphology of the
stellar and ionised gas component are similar although the stellar component
presents a  more spread distribution than that of the ionised gas. In
the following, we describe briefly the identified regions:

   \begin{figure*}
   \centering
\includegraphics[angle=0,scale=1.00, clip=,bbllx=67, bblly=239, bburx=545,
bbury=787]{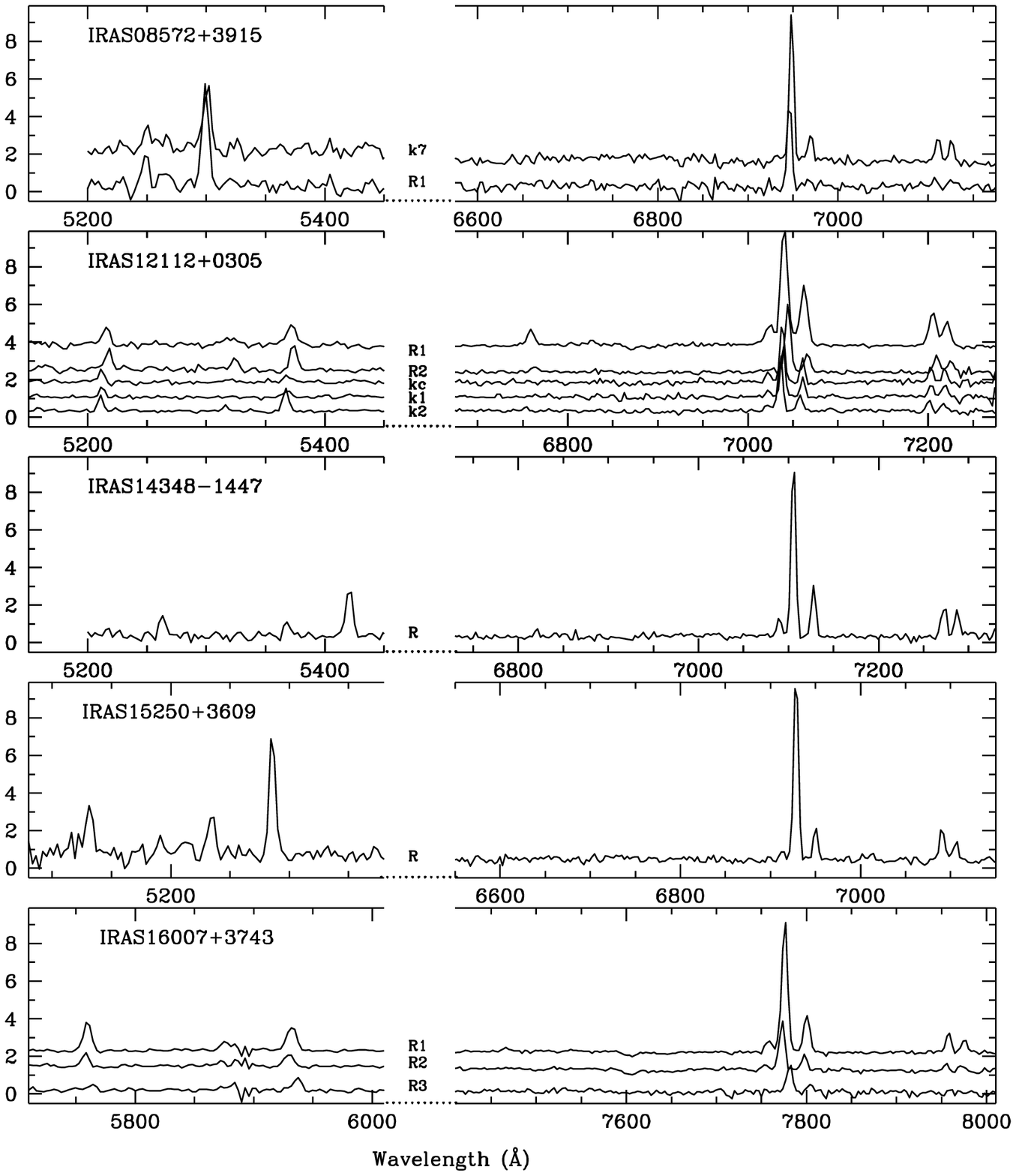}
   \caption{Representative observed spectra of the main extranuclear
     line-emitting 
  regions. Note that the horizontal axis does not represent a continuous range
  in wavelength, but two subsets of the full INTEGRAL
  spectrum corresponding to
  H$\beta$--\textsc{[O\,iii]}$\lambda\lambda$4959,5007 and 
  \textsc{[O\,i]}$\lambda$6300 --
  H$\alpha$+\textsc{[N\,ii]}$\lambda\lambda$6548,6586 --
  \textsc{[S\,ii]}$\lambda\lambda$6717,6731  
  spectral regions, respectively. Flux units are arbitrary and for the shake
  of clarity, offsets have been applied to the
  spectra. \label{espectros}  
}
    \end{figure*}

\begin{description}

\item[\textbf{IRAS~08572+3915:}]  Two condensations have been identified
  here: k7 at $\sim 6^{\prime\prime}$ from the northern
  galaxy toward the northeast, near its tidal tail \citep{arr00} and
  R1 \citep[k4+5 in the notation of ][]{arr00} in the southern tidal
  tail and at a distance of $\sim 5\farcs7$ from the 
  southern nucleus. Here k4+5 will be treated as a single region as
  we were not 
  able to close independent contours for them in the WFPC2 image.

\item[\textbf{IRAS~12112+0305:}] This system shows several regions of
  interest. The brightest emission line region (R1) is not associated with
  any of the nuclei of the system, but with a low surface brightness
  region in the WFPC2 image towards the east of the system
  \citep{col00}. This image also shows an arc-like condensation located
  3$^{\prime\prime}$ north of the southern nucleus at P.A.~$\sim
  20^{\mathrm{o}}$ (R2) which 
  is relatively bright in the H$\alpha$ emission line map. Finally,
  the southern tidal tail shows a series of condensations which
  correspond either to peaks in the emission line map (k1 and k2) or
  in the continuum map (kc).

\item[\textbf{IRAS~14348$-$1447:}] The most interesting region in this system
  has already been identified by \citet{mih98} as a peak of H$\alpha$
  emission towards the southeast of the system (R). In addition, there are
  other two regions which should be mentioned here. On the one hand,
  the peak of 
  H$\alpha$ emission in the northern galaxy is not associated with its
  nucleus but with four 
  knots located at the base of its tidal tail. Similarly to
  IRAS~15206+3342, the individual knots cannot be resolved in the ground based
  data and its analysis will not be considered here. On the other hand, the
  WFPC2 image shows a faint condensation at $\sim$4\farcs0 and P.A.$\sim
  70^\mathrm{o}$ from the southern nucleus whose analysis will also not be
  considered here due to the elevated flux contamination 
  from the southern galaxy in the INTEGRAL data.

\item[\textbf{IRAS~15250+3609:}] This galaxy presents a low surface brightness
  region in the F814W band at $\sim$7\farcs0 from the nucleus towards
  the southwest (R), which is however relatively bright in the line
  emission maps \citep{mon04}. HST infrared images show a series of
  knots in the main body of the galaxy probably associated with
  globular clusters \citep{sco00}. Although they can be responsible for
  some of the extended emission observed in the ground based data,
  they cannot be identified as individual sources and their analysis
  is not considered here. 

\item[\textbf{IRAS~16007+3743:}] The morphology
of IRAS~16007+3743 could indicate that this system is a multiple merger
\citep{cui01,bor00}. However, kinematical information derived from INTEGRAL
(i.e. velocity field and velocity dispersion maps) made us classify this
system as a system of two 
spiral galaxies in interaction \citep{gar05}. Three regions have been
identified: R1 and R2 
  seen in the H$\alpha$ map as two bright condensation along the
  northern tidal tail and R3, a region at $\sim$3\farcs0 towards
  the south of the system  apparently not associated with any of the
  tidal tails.
\end{description}


To have an idea of the data quality, the observed spectra for the
regions under study
are shown in Figure \ref{espectros}. From blue to red,
H$\beta$, \textsc{[O\,iii]}$\lambda\lambda$4959,5007,
\textsc{[O\,i]}$\lambda$6300,
H$\alpha$+\textsc{[N\,ii]}$\lambda\lambda$6548,6584 and
\textsc{[S\,ii]}$\lambda\lambda$6717,6731 are detected in most of the regions
with enough signal-to-noise.

   \begin{table*}
\centering
      \caption[]{Properties of the Star-Forming Regions (I): Emission line
        ratios. \label{props_1}}
              \begin{tabular}{cccccccccccccc}
            \hline
            \noalign{\smallskip}
Region &
$\Delta\alpha^{\mathrm{a}}$ &
$\Delta\delta^{\mathrm{a}}$ &
$E(B-V)^{\mathrm{b}}$ &
\textsc{[O\,iii]}/H$\beta^{\mathrm{b,c}}$ &
\textsc{[O\,i]}/H$\alpha^{\mathrm{b,c}}$ &
\textsc{[N\,ii]}/H$\alpha^{\mathrm{b,c}}$ &
\textsc{[S\,ii]}/H$\alpha^{\mathrm{b,c}}$ &
$12 + \log (\mathrm{O/H})^{\mathrm{d}}$ &
$12 + \log (\mathrm{O/H})^{\mathrm{e}}$ 
\\  
 &
($^{\prime\prime}$) &
($^{\prime\prime}$) &
 &
 &
 &
 &
 &
(\textsc{[O\,iii]}/H$\beta_{\mathrm{upper}}$) & 
(N2) &
\\
            \noalign{\smallskip}
            \hline
            \noalign{\smallskip}
\hline
\multicolumn{9}{c}{IRAS~08572+3915$^{\mathrm{f}}$}\\
\hline
R1$^{\mathrm{g}}$ & 5.2 & $-$7.4 & \ldots & 0.24 & \ldots & $-$0.85 & \ldots &
8.49 & 8.50\\
k7 & 5.3 &  2.2 & \ldots & 0.54 & \ldots & $-$0.82 & \ldots &
8.29 & 8.52\\
\hline
\multicolumn{9}{c}{IRAS~12112+0305}\\
\hline
R1 & 2.7    & $-$1.5  & 0.25 &    0.22 & $-$1.24 & $-$0.66 & $-$0.58 &
8.50 & 8.64\\ 
R2 & $-$1.5 &    0.6  & 1.07 &    0.01 & $-$1.02 & $-$0.41 & $-$0.43 &
8.65 & 8.82\\
kc & $-$1.8 & $-$7.4  & 0.47 &    0.16 & \ldots  & $-$0.64 & $-$0.99 &
8.55 & 8.65\\
k1 & $-$2.8 & $-$7.2  & 0.66 & $-$0.12 & \ldots  & $-$0.49 & $-$0.38 &
8.74 & 8.76\\
k2 & 0.7    & $-$7.7  & 0.45 &    0.08 & \ldots  & $-$0.42 & $-$0.39 &
8.60 & 8.81\\
\hline
\multicolumn{9}{c}{IRAS~14348$-$1447}\\
\hline
R & 5.1    & $-$1.7  & 1.12   &    0.26 & $-$1.36 & $-$0.55 & $-$0.49 &
8.48 & 8.72\\
\hline
\multicolumn{9}{c}{IRAS~15250+3609}\\
\hline
R  & $-$5.0 & $-$4.8  & 0.25   & 0.26    & \ldots & $-$0.78 & $-$0.60 &
8.48 & 8.55\\
\hline
\multicolumn{9}{c}{IRAS~16007+3743}\\
\hline
R1 & 0.9    &    5.4  & 0.55 & 0.13    & $-$1.05 & $-$0.47 & $-$0.67 &
8.59 & 8.78\\
R2 & 0.7    &    2.8  & 0.40 & $-$0.02 & \ldots & $-$0.55 & $-$0.67 &
8.65 & 8.71\\
R3 & 2.2    & $-$1.9  & 0.75 & 0.49    & \ldots  & $-$0.55 & \ldots &
8.59 & 8.75\\
            \noalign{\smallskip}
            \hline
         \end{tabular}
\begin{list}{}{}
\item[$^{\mathrm{a}}$] Relative positions of the regions. We took the northern
  nucleus as reference in IRAS~08572+3915 and IRAS~12112+0305; the southern
  one in IRAS~14348$-$1447 and the western one in IRAS~16007+3743.
\item[$^{\mathrm{b}}$] Calculated as the average value in a 0\farcs45-radius
  aperture centred in the region of interest.
\item[$^{\mathrm{c}}$] Extinction corrected data. We assumed a foreground
  screen model, and used the reddening curve of \citet{whi58} parametrised as
  explained in 
\cite{mil72} and R=3.1 \citep{rie85}, except for regions R1 and k7 of
IRAS~08572+3915 where no extinction measurements are available. Typical errors
for line ratios are $\lsim$0.2 dex.
\item[$^{\mathrm{d}}$] Derived metallicities using the upper branch of the
  empirical diagram of \citet{edm84}. 
\item[$^{\mathrm{e}}$] Derived metallicities using the N2 calibrator of
  \citet{den02}. 
\item[$^{\mathrm{f}}$] Measured values taken from \citet{arr00}.
\item[$^{\mathrm{g}}$] This corresponds to knots k4 and k5 in the notation of
  \citet{arr00}. Their relative positions are:  (5.8,$-$7.8) and (4.7,$-$7.0).

\end{list}
   \end{table*}

\subsection{Characterisation of the external star-forming regions}

\subsubsection{Line ratios}

   \begin{figure*}
   \centering
\includegraphics[angle=0,scale=0.9, clip=,bbllx=32, bblly=325, bburx=572,
bbury=534]{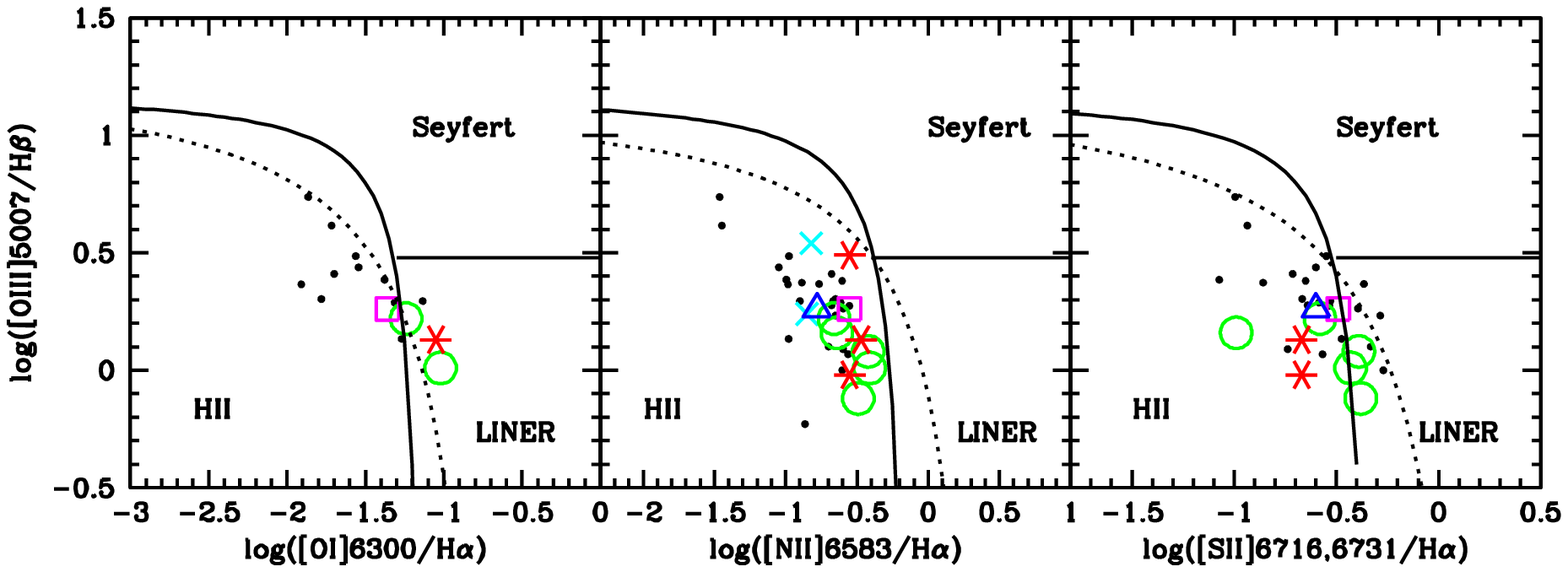}
      \caption{Position of the different regions in the diagnostic diagrams
        proposed 
by \citet{vei87}. The solid lines show the empirical borders found by these
authors between the different ionisation types while the dashed lines show the
theoretical borders proposed by \citet{kew01} to delimit the area where the
line ratios can be explained by star formation. The used colour/symbol code is
the following: IRAS~08572+3915, cyan/cross; IRAS~12112+0305, green/circle;
IRAS~14348$-$1447, magenta/square; IRAS~15250+3609, blue/triangle;
IRAS~16007+3743, red/asterisk. Typical errors are
$\sim$ 0.2 dex. The measured ratios in other
TDG candidates are shown for comparison as black dots
\citep{duc98,tem03,lop04}. 
\label{diagdiag}
}
   \end{figure*}

Measured line ratios are shown in Table \ref{props_1} and
Figure \ref{diagdiag},  which contain the classical diagnostic diagrams of
\citet{vei87}. The regions present line ratios typical of \textsc{H\,ii}
regions. Only by means of the [O\,\textsc{i}]$\lambda$6300/H$\alpha$ line
ratio some of them could marginally be classified as LINER. Star formation is
the most plausible mechanism to explain the observed line 
ratios i.e. the line ratio of all the regions analysed here are well within
the limits proposed by \citet{kew01} for ionisation by star formation.



In general, line ratios are comparable to those found in TDGs
candidates \citep[e.g.][]{duc98,tem03,lop04}, except for their
[O\,\textsc{i}]$\lambda$6300/H$\alpha$ which is smaller by $\sim$0.3
dex in those cases where 
it could be measured.  However, the poor statistics (i.e. the
[O\,\textsc{i}]$\lambda$6300 line could only be measured for four regions)
makes difficult to assess the significance of this result.

\subsubsection{Metallicities \label{metalicidades}}

One observational characteristic that helps to establish the tidal origin of a
certain dwarf galaxy is its high metallicity, which suggests that TDGs are made
up from processed material \citep{duc98}. 
Abundances are usually estimated using empirical methods based on
the intensities of certain optical lines like the widely used
method of the $R_{23}$ calibrator \citep[i.e.][]{tor89} and the
$S_{23}$ indicator \citep{vil96}. However, both of them involve
emission lines which are not available within our spectral range.
Nevertheless, it is possible to estimate
the metallicity using the N2 calibrator proposed by
\citet{den02} -- based on the ratio between the \textsc{[N\,ii]}$\lambda$6584
and H$\alpha$ emission lines --, and the empirical diagrams of \citet{edm84},
-- based on the \textsc{[O\,iii]}$\lambda\lambda$4959,5007 to H$\beta$ ratio --
parametrised as explained in \citet{duc98}. The predicted metallicity values
expected from these indicators are shown in the last two columns of Table
\ref{props_1}. 
Both indicators predict similar values for the metallicity (average of
$12 + \log \mathrm{O/H} = 8.70$ for the N2 calibrator against 8.57
for the one involving the oxygen lines). Giving that typical
uncertainties for the line ratios are about 0.2~dex (which translates into
$\sim$0.3 for the metallicity), differences between both indicators are
not significative.  
This relatively high metallicities are about a factor 2 larger than
those derived 
for other TDGs candidates \citep[$\sim$8.35,][]{duc98,wei03} and about
half the solar metallicity.

   \begin{table*}
\centering
      \caption[]{Properties of the Star-Forming Regions (II): H$\alpha$
        related observables, I magnitudes, sizes and
        distances. \label{props_2}} 
              \begin{tabular}{cccccccccccccc}
            \hline
            \noalign{\smallskip}
Region &
EW(H$\alpha$)$^{\mathrm{a}}$ &
F$_{obs}$(H$\alpha$)$^{\mathrm{b}}$ &
L(H$\alpha$)$^{\mathrm{c}}$ &
m$_I^{\mathrm{d}}$ &
M$_I^{\mathrm{e}}$ &
r$_{\mathrm{equ}}$ &
r$_{\mathrm{eff}}$ &
$D_{\mathrm{CM}}^{\mathrm{f}}$ &
$D_{\mathrm{near}}^{\mathrm{g}}$ 
\\  
 & 
(\AA) &
(10$^{-16}$~erg~s$^{-1}$~cm$^{-2}$) &
(10$^{40}$~erg~s$^{-1}$)  &
 &
 &
(pc) &
(pc) &
(kpc)  &
(kpc)  
\\
            \noalign{\smallskip}
            \hline
            \noalign{\smallskip}
\hline
\multicolumn{10}{c}{IRAS~08572+3915$^{\mathrm{h}}$}\\
\hline
R1 & 78 & 7.0 & 0.6 & 23.78 & $-$13.30 & 373 & 121 & 8.8 & 4.2\\
k7 & 245 & 4.0 & 0.3 & 22.62 & $-$14.46 & 239 & 128 & 6.4 & 6.4\\
\hline
\multicolumn{10}{c}{IRAS~12112+0305}\\
\hline
R1 & 257 & 31.8 & 7.4 & 21.12 & $-$16.86 & 373 & 224 & 5.0 & 4.3\\
R2 & 79  & 37.4 & 57.6& 19.05 & $-$20.14 & 698 & 494 & 2.4 & 2.2\\
kc & 34  & 2.7 & 1.0 & 21.15 & $-$17.16 & 451 & 311 & 9.0 & 6.9\\
k1 & 46  & 3.5 & 2.1 & 21.43 & $-$17.16 & 383 & 150 & 9.0 & 7.1\\
k2 & 71  & 3.4 & 1.2 & 22.02 & $-$16.26 & 378 & 249 & 9.0 & 8.4\\
\hline
\multicolumn{10}{c}{IRAS~14348$-$1447}\\
\hline
R & 194 & 15.5 & 34.1 & 20.40 & $-$19.13 & 1082 & 546 & 8.2 & 6.0\\
\hline
\multicolumn{10}{c}{IRAS~15250+3609}\\
\hline
R  & 141 & 6.5 & 0.6 & 20.73 & $-$16.36 & 891 & 420 & 6.9 & 6.9\\
\hline
\multicolumn{10}{c}{IRAS~16007+3743}\\
\hline
R1 & 40  & 12.0 & 57.1  & 20.16 & $-$20.63 & 1151 & 828 & 17.1 & 16.9\\
R2 & 234 & 35.5 & 189.5 & 19.80 & $-$21.07 & 1375 & 884 & 9.4 & 8.9\\
R3 & 80  & 5.85 &    8.8 & 21.83 & $-$19.23 & 1562 & 851 & 5.8 & 9.0\\
            \noalign{\smallskip}
            \hline
         \end{tabular}
\begin{list}{}{}
\item[$^{\mathrm{a}}$] Calculated as the average value in an 0\farcs45-radius
  aperture centred in the region of interest.
\item[$^{\mathrm{b}}$] As a compromise between collecting all the flux of the
  region and avoiding the contamination from the neighbouring zones of the
  systems, flux was measured in a 1\farcs0-radius aperture in all regions of
  IRAS~08572+3915, IRAS~14348$-$1447, IRAS 15250+3609 and R1 and R2 of
  IRAS~12112+0305 while for those in the tidal tail of this system and
  those in IRAS~16007+3743, a 0\farcs5-radius aperture was used.
\item[$^{\mathrm{c}}$] Extinction corrected H$\alpha$ luminosity except for
  regions R1 and k7 of IRAS~08572+3915 where no extinction measurements are
  available. 
\item[$^{\mathrm{d}}$] Observed I magnitude, measured within the area
  enclosed by the biggest closed 
  contour centred in the emission peak (see explanation in text).
\item[$^{\mathrm{e}}$] Extinction corrected absolute I magnitude.
\item[$^{\mathrm{f}}$] Distance to the mass centre of the system.
\item[$^{\mathrm{g}}$] Distance to the nearest galaxy.
\item[$^{\mathrm{h}}$] H$\alpha$ equivalent widths and fluxes taken from
  \citet{arr00}. 

\end{list}
   \end{table*}

\subsubsection{Sizes \label{tamagnos}}

Sizes were estimated from the WFPC2/F814W images. The irregular shape of the
regions as well as some possible contamination 
from other structures within the system (i.e. other regions, tails or the
parent galaxy itself) made difficult to define their limits. In order to
proceed in a systematic way and compare with the sizes of other extragalactic
objects, we defined the 
size of a given region as \emph{the area encircled in the largest
closed contour centred in the emission peak associated with that
region}. This allows us to define an equivalent 
radius as $r_{\mathrm{equ}} = \sqrt{\mathrm{Area}/\pi}$ which gives an
estimate of the \emph{total} size of the region. An effective radius
was also measured
as the one which contains half of the flux within this
area (see columns 7 and 8 of table \ref{props_2}). They can be use for
comparison with the characteristic radii derived for other
objects. The ratio between the effective  and equivalent radii  
($0.3 - 0.7$) gives and idea of the compactness of a certain condensation. 

Equivalent radii range from a few hundreds pcs to $\sim 1.5$~kpc (mean of
750~pc). In general, when comparing with the sizes of \textsc{H\,ii} regions
(which have typical radii of $\sim100-900$~pc), the present regions are
similar to the largest Giant \textsc{H\,ii} regions \citep{ken84,may94}. 

All the regions under study have effective radii (mean $\sim 430$~pc)
comparable to dwarf galaxies located in the Local Group 
\citep[$r \sim$ 0.3~kpc,][]{mat98}, to the so-called Blue Compact
Dwarf galaxies, with effective radii  $0.2 \lsim r_{\mathrm{eff}}
\lsim 1.8$~kpc \citep[e.g.][]{mar97,cai03} or to some already detected
TDG \citep[e.g.][]{duc98}. 

%

\subsubsection{H$\alpha$ equivalent widths and luminosities \label{sec_lha}}

   \begin{figure}
   \centering
\includegraphics[angle=0,scale=0.5, clip=,bbllx=20, bblly=210, bburx=526,
bbury=643]{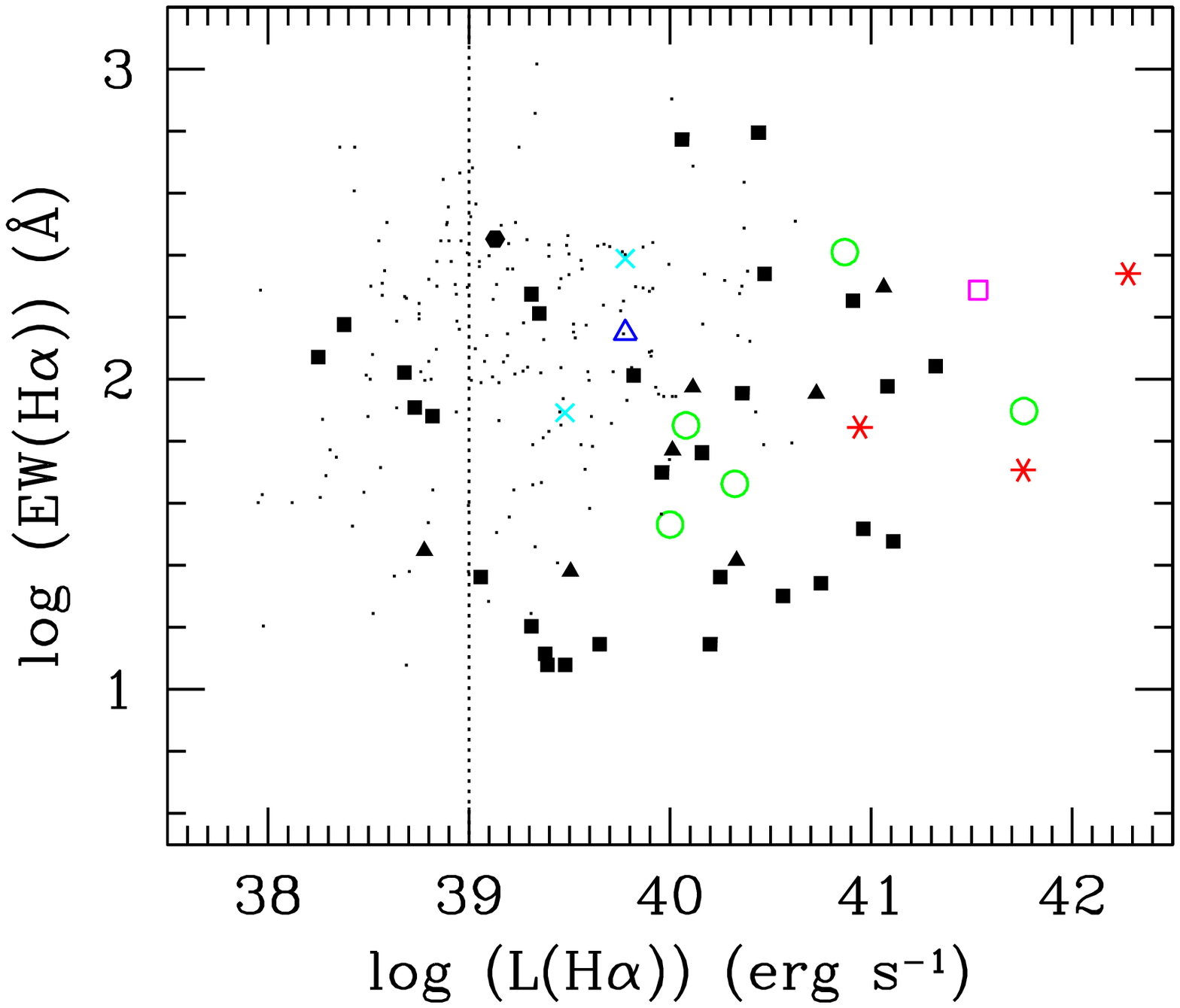}
      \caption{H$\alpha$ luminosities vs. equivalent widths. Data from the
    present study follow the same colour/symbol code as in Figure
 \ref{diagdiag}. We also include for comparison 
 measurements for the extragalactic \textsc{H\,ii} regions of \citet{may94}
 (dots, not extinction corrected) as well as for other
 extranuclear star forming regions which have been analysed as possible TDG
 candidates (solid symbols). Those of  \citet{igl01} (squares) and
 \citet{mun04} (hexagon) are not extinction corrected while those of
 \citet{tem03} (triangles) have been corrected for extinction. The vertical
    dashed line shows the luminosity limit for TDG candidates used in
    \citet{igl01}. \label{fig_lhavsewha}
}
   \end{figure}
%

   \begin{figure}
   \centering
\includegraphics[angle=0,scale=0.5, clip=,bbllx=20, bblly=220, bburx=526,
bbury=643]{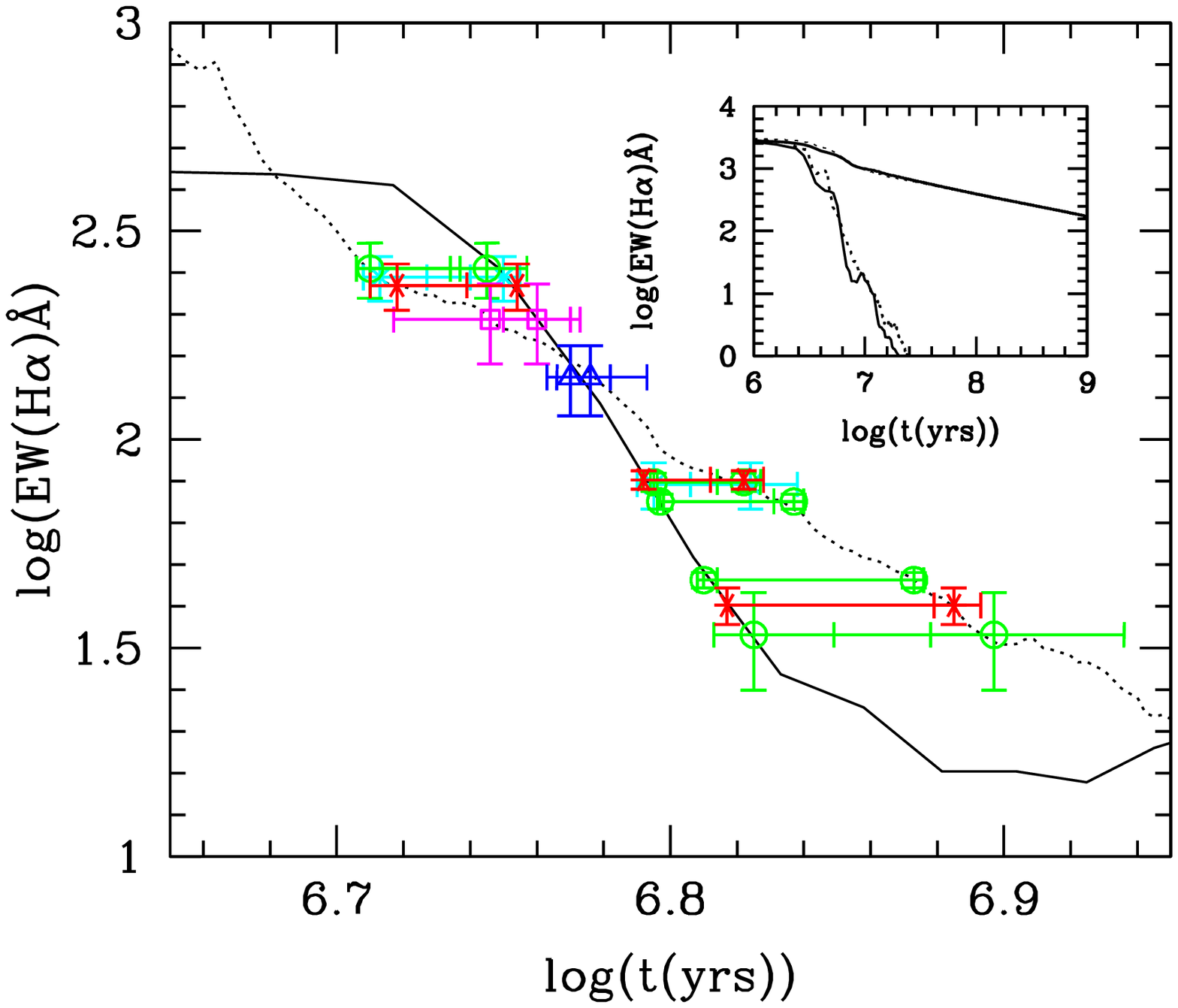}
      \caption{Measured H$\alpha$ equivalent widths in comparison to two SB99
  instantaneous 
  bursts of $Z = 0.020$ (continuum line) and $Z = 0.008$ (dashed
  line). Colour/symbol 
  code is the same as in figure \ref{diagdiag}. The
  predicted ages range from 5.5 to 7.1~Myr if the model with solar metallicity
  is assumed or from 5.1 to 7.8~Myr if the one for lower metallicity is used.
  The small inset shows the covered range in equivalent width for both
  continuous and instantaneous STARBURSTS99 models (thin and thick lines
  respectively). Note the elevated
  equivalent widths predicted for the continuous star formation models, which
  are not reached 
  by any of the regions analysed here. \label{fig_ewha} 
}
   \end{figure}
%

H$\alpha$ equivalent width and luminosity distributions are shown in Figure
\ref{fig_lhavsewha} together with the values measured for some TDGs as well as
\textsc{H\,ii} regions taken from the literature. 
The relatively high H$\alpha$ equivalent widths derived (i.e. $34 - 257$~\AA),
indicative of 
the existence of a very young stellar population, are comparable to
those measured for TDG candidates
(e.g. $12 - 623$~\AA{} for \citealt{igl01} or $24 - 198$~\AA{} for
\citealt{tem03}) 
and for extragalactic \textsc{H\,ii} regions in general \citep{may94}.
Regarding the H$\alpha$ luminosities, once the extinction
effects have been taken into account, the selected regions are more
luminous than the Giant \textsc{H\,ii} Regions \citep{may94,ken84}. In 
general, these luminosities are in better agreement with those found for
TDG candidates \citep{duc98,igl01}. The most suitable sample for
comparison with our
candidates is the one of \citet{tem03} which takes into account extinction
effects. With the exception of three very bright regions (see discussion
below), the H$\alpha$ luminosities are comparable to those of Temporin
et al. (mean $\sim 3\times 10^{40}$ erg s$^{-1}$ for their sample against
$\sim 6\times 10^{40}$ erg s$^{-1}$ for ours). 

The extinction corrected luminosity of three regions are well above
$5\times10^{41}$~erg~s$^{-1}$ which is more than one order of 
magnitude larger than the luminosities found in typical TDGs. Two of them
belong to 
IRAS~16007+3743 which is at z=0.185 and has a factor of $2-3$ lower
linear resolution 
than the rest of the sample. One possibility is that emission from these two
particular regions was due to a wider area that encompasses more than one
condensation.  However, a resolution effect cannot be used as explanation for
the high luminosity of the third region (R2 of IRAS~12112+0305, scale
$\sim$1.4 kpc arcsec$^{-1}$).
Alternatively, it may be possible that the relatively large
amount of gas involved in ULIRGs in general, and the violent event suggested
by the morphology of IRAS~16007+3743 and IRAS~12112+0305 could lead to the
large luminous TDGs.  

It is widely known that H$\alpha$ emission can be used to constrain the
properties of recent episodes of star formation.  An upper limit to
the age can be estimated by using
its equivalent width, as it decreases with time (see inset in Figure
\ref{fig_ewha}). We have used STARBURST99 models \citep{lei99} to estimate
ages for the bursts in the extranuclear regions.
In view of the estimated metallicities in section \ref{metalicidades}, we
chose two different spectral synthesis models, 
both for an instantaneous burst with a Salpeter IMF, and upper mass limit of
125~M$_\odot$ but with two different metallicities: $Z = 0.020$ and $Z =
0.008$. Models with continuous star formation cannot
explain the observed equivalent widths as the newly born massive stars
are able to maintain too elevated values of the equivalent width in comparison
to the observed ones even for ages greater than 100~Myrs. 
The estimated ages for the bursts  are shown in Figure \ref{fig_ewha} and
second column of table \ref{props_3}. They range between $5 - 8$~Myr, depending
on the assumed model, confirming the presence of a young population.

Summarising, equivalent widths of the analysed regions are similar to those
derived for both extragalactic \textsc{H\,ii} regions and TDG
candidates and they are
typical of young burst of star formation. The H$\alpha$ luminosity place these
regions more in the group of TDG candidates than in that of extragalactic
\textsc{H\,ii} regions.

   \begin{table*}
\centering
      \caption[]{Properties of the Star-Forming Regions (III): Derived
        characteristics of the stellar populations and dynamical
        parameters. \label{props_3}} 
              \begin{tabular}{cccccccccccccc}
            \hline
            \noalign{\smallskip}
Region  &
Age$^{\mathrm{a}}$ &
$m_{y.b.}^{\mathrm{a}}$ &
$\sigma^{\mathrm{b}}$ &
$M_{\mathrm{dyn}}$ &
$M^{\mathrm{A}}_{\mathrm{tid}}$$^{\mathrm{c}}$ &
$M^{\mathrm{B}}_{\mathrm{tid}}$$^{\mathrm{d}}$ &
$v_{\mathrm{rel}}$ &
$v_{\mathrm{esc}}$ &
$M_{\mathrm{dyn}} / M_{\mathrm{tid}}$
\\  
 & 
($10^6$~Myr) &
($10^6$~M$_\odot$) &
(km~s$^{-1}$) &
(10$^8$~M$_\odot$) &
(10$^8$~M$_\odot$) &
(10$^8$~M$_\odot$) &
(km~s$^{-1}$)&
(km~s$^{-1}$)
\\
            \noalign{\smallskip}
            \hline
            \noalign{\smallskip}
\hline
\multicolumn{10}{c}{IRAS~08572+3915}\\
\hline
R1 & 5.4 & 1.9 & \ldots & \ldots &  0.04 & 0.09 & 196 & 133 & \ldots\\
k7 & 6.4 & 2.2 & 47 & 5.9 &  0.03 & 0.02 &  41 & 156 & 196\\
\hline
\multicolumn{10}{c}{IRAS~12112+0305}\\
\hline
R1 & 5.3 & 22.8  & 60 & 16.8 &  1.22 & 1.10 & 75     & 410 & 14\\
R2 & 6.4 & 420.6 & 56 & 32.4 &  72.32 & 53.65 & $-$280 & 592 & 0.4\\
kc & 7.3 & 11.5  & \ldots & \ldots  & 0.37 & 0.32 & $-$338 & 306 & \ldots\\
k1 & 6.9 & 19.6  & \ldots & \ldots  & 0.23 & 0.21 & $-$207 & 306 & \ldots\\
k2 & 6.6 & 9.4   & \ldots & \ldots  & 0.18 & 0.11 & $-$343 & 306 & \ldots\\
\hline
\multicolumn{10}{c}{IRAS~14348$-$1447}\\
\hline
R & 5.7 & 136.6 & 50  & 28.5  & 10.6 & 14.78 & 144   & 402 & 1.9\\
\hline
\multicolumn{10}{c}{IRAS~15250+3609}\\
\hline
R  &  6.5 & 2.8 & 50 & 21.9 & 3.62 & 3.62 & 170 & 264 & 6.0\\
\hline
\multicolumn{10}{c}{IRAS~16007+3743}\\
\hline
R1 & 7.1 & 581.6  & 61 & 64.4  & 1.7 & 1.1 & $-$41 & 302 & 38\\
R2 & 5.4 & 678.3  & 76 & 106.8 & 17.7 & 112.39 & 92 & 408 & 0.9\\
R3 & 6.4 & 62.8   & 85 & 128.5 & 106.65 & 12.58 & 338 & 519 & 1.2\\
           \noalign{\smallskip}
            \hline
         \end{tabular}
\begin{list}{}{}
\item[$^{\mathrm{a}}$] The age and mass in young stars obtained as the average
  from the predictions for the low and high metallicity models.
\item[$^{\mathrm{b}}$] Calculated as the average value in an 0\farcs45-radius
  aperture centred in the region of interest.
\item[$^{\mathrm{c}}$] Tidal mass assuming the potential is created by a point
  mass in the mass centre of the system.
\item[$^{\mathrm{d}}$] Tidal mass assuming the potential is created by the
  nearest galaxy. 

\end{list}
   \end{table*}

\subsubsection{I magnitudes}

   \begin{figure}
   \centering
\includegraphics[angle=0,scale=0.5, clip=,bbllx=35, bblly=185, bburx=535,
bbury=645]{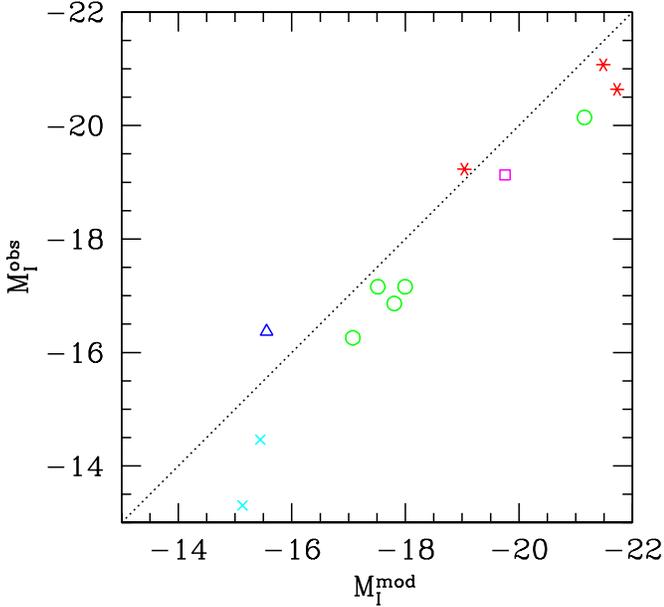}
      \caption{Comparison between measured and predicted I
        magnitudes. Colour/symbol code is the same as in Figure
        \ref{diagdiag}. The dotted line represents slope one.
        Predicted magnitudes were obtained as an average
        of the predicted magnitudes for the $Z = 0.020$ and $Z =
        0.008$ instantaneous burst models.  \label{compa_magi}
}
   \end{figure}
%

We have measured the I magnitudes within the apertures defined in
section \ref{tamagnos} using the task POLYPHOT under the IRAF
environment.
Typical errors in the measurements where $\sim 0.1$, with the
exception of R1 in IRAS~08572+3915 that was $\sim 0.4$.
The comparison between these magnitudes and those derived from STARBURST99 are
shown in Figure \ref{compa_magi} where plotted magnitudes were
obtained as an average of the predicted magnitudes for the $Z = 0.020$ and $Z =
0.008$ instantaneous burst models.
The trend is good although the predicted values from
  extinction corrected H$\alpha$ luminosities overestimate the
  observed magnitudes by $\sim$0.9 mag on average. These differences
  can be explained by both, the uncertainties associated with the
  selection of the aperture in the different bands and the  existence of
other stellar populations either from a previous burst during the merging
process or from the original population in the merging galaxies, as found in
some TDG candidates \citep[e.g.][]{lop04}.
The presence of such a older population
would cause an underestimation of the equivalent width and thus,
overestimation of the age of the burst. Younger ages for the bursts
translates into a decrease of the predicted I magnitudes of $\sim 0.5 - 1.0$
mag according to the models.

\subsubsection{Distances}

The two last columns in Table \ref{props_2} show the projected distances
between the studied regions and the mass
centre of the system, as well as the closer galaxy. Typical measured
distances are $\sim$8~kpc with a maximum of $\sim$17~kpc. 
However, these objects were not selected based on a particular
  favourable geometry. Indeed, their geometries are 
very complex and the influence of projection effects -- which
should be present -- is particularly difficult to quantify. These
distances have to be seen as lower limits to the actual ones.
These are relatively close distances to the parent galaxy when compared with
other condensations in H$\alpha$ identified as
TDGs  \citep[e.g.][]{duc98} although there are already some
  examples of  TDG candidates at such distances (see for instance
  TD44a in \citealt{igl01} or objects A2 and E in \citealt{amr04}).

\section{Discussion: What are the chances of survival of the analysed
  regions as TDGs?}

Most of the observational properties of the selected regions
(i.e. H$\alpha$ equivalent widths and luminosities, metallicities,
ionisation states and radii) derived in section \ref{resultados} place these
regions in the locus of the most 
luminous extragalactic Giant \textsc{H\,ii} Regions and are consistent
with those expected for TDGs or TDG progenitors.
In the next section, we will use the kinematical information
provided by INTEGRAL which, together with the evolutive state of the system,
will allow us to estimate the likeliness of survival of these candidates.

As TDGs are stable entities with their own dynamic, the best definition
for this kind of objects is the one  proposed by \citet{duc00} who
identify a TDG as that object that constitutes a self-gravitating entity
and that was made up from the debris of a galaxy interaction.
To evaluate the chance of survival of TDG candidates, two basic questions need
to be answered:
i) Is the candidate massive enough to survive to
its internal movements?  ii) Is the candidate massive enough to survive to the
gravitational forces exerted by the parent galaxy?
Depending on the available observables, several criteria
have been used in the literature in the past to classify 
a certain condensation as a TDG. In the following we apply them (when
possible) to our star-forming regions. 

\subsection{Stability against internal motions}

\citet{igl01} established a  luminosity criterion ($L(\mathrm{H}\alpha) >
10^{39}$~erg~s$^{-1}$) which should fulfil systems stable to internal
motions. As it has been mentioned above (section \ref{sec_lha}) all the
selected candidates meet well this criterion  (see also Figure 4), especially
two very luminous regions of IRAS 16007+3743 and R2 in IRAS~12112+0305.  

   \begin{figure}
   \centering
\includegraphics[angle=0,scale=0.5, clip=,bbllx=30, bblly=205, bburx=535,
bbury=645]{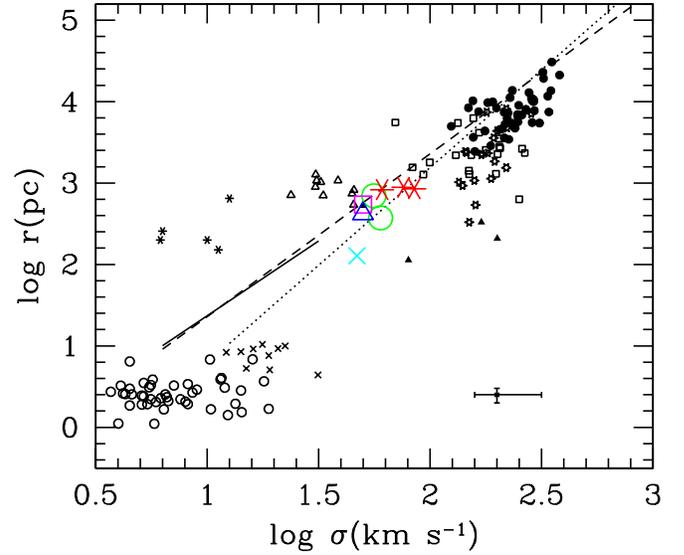}
      \caption{Velocity dispersion vs. estimated effective
        radius. The colour/symbol code is the same as in figure 
  \ref{diagdiag}. The typical size for errors is
  shown in the lower right corner. The continuous line 
  represents the fit for extragalactic \textsc{H\,ii} regions obtained by
  \citet{ter81} while the dashed and dotted lines show their fit for
  elliptical galaxies, and globular cluster + elliptical galaxies
  respectively.   Small symbols represent different samples
  of dynamically hot systems: open circles, galactic globular clusters
  \citep{tra93,pry93}; crosses, massive globular clusters in NGC~5128
  \citep{mar04}; open triangles, dwarf elliptical galaxies \citep{geh03}; open
  squares, intermediate ellipticals; solid circles, giant 
  ellipticals;  solid triangles, compact ellipticals; asterisks, dwarf 
  ellipticals; and stars, bulges
  \citep[all of them from][]{ben92}.\label{fig_rsigma}  
} 
   \end{figure}
%

One of the best test to check if a certain detected
  condensation constitutes 
a self-gravitating entity is detecting velocity gradients that could
indicate the existence of independent rotation
\citep[e.g.][]{wei02,bou04,men01}. However our data lacks the
  necessary angular resolution to resolve any velocity field across the
  extranuclear condensations.
  Typical size of the detected regions is roughly equivalent to 1-3
  SB2 fibres, which doesn't allow to derive the 
  individual velocity field of the regions under study.

Another method that can be used to establish whether or not
the TDG candidates are stable, self-gravitating entities is to study
their location in the radius-velocity dispersion correlation measured
for ellipticals and globular clusters.
With this purpose, we have plotted  the classical fits in the $r_{eq}
- \sigma$ plane for elliptical galaxies, globular clusters and \textsc{H\,ii}
regions of \citet{ter81} together with our data in
Figure \ref{fig_rsigma}. For the present INTEGRAL 
  configuration the spectral resolution is $\sigma \sim$80~km s$^{-1}$. We
  considered that we were able to determine a reliable measurement of the
  velocity dispersion when this was greater that half the spectral resolution.
We have also overplotted the position of recent samples of hot
dynamically system 
such as galactic globular clusters  \citep{tra93,pry93}, massive globular
clusters in in NGC5128 \citep{mar04}, dwarf elliptical galaxies in the Virgo
cluster \cite{geh03},  and the sample of dynamically hot systems
of \citet{ben92}. Based on their position in the radius-velocity dispersion
diagram, the selected TDG candidates should be self-gravitating 
entities. Moreover, they are in a region close to other dwarf galaxies
\citep[compare in particular with the sample of ][]{geh03}, and far from the
region occupied by globular clusters.

\subsection{Stability against forces from the parent galaxy}

One way to see whether a condensation is stable against the
gravitational potential of the parent galaxy/system is estimating its
tidal mass \citep{bin87,men01} which is defined as:

\begin{equation}
M_{tid} = 3 M \left(\frac{R}{D}\right)^3
\end{equation}

where $M$ is the mass of the parent galaxy, $R$ is the radius of the
TDG candidate (here estimated as $r_{equ}$) and $D$ is the distance to the
parent galaxy. If the tidal mass is smaller 
than the mass for the TDG candidate,
then the candidate is stable against the forces exercised by the
parent galaxy. This expression is valid in those cases where the size of a
certain region is small in comparison with the distance to the parent galaxy
\citep{bin87}. Typical values for the ratio \emph{size / projected distance}
(which is an upper limit for the real \emph{size / distance}) for
our condensations range between 0.04 (k2 in IRAS~12112+0305) and 0.32 (R2 in
IRAS~12112+0305), small enough to try to asses their stability against forces
from the parent galaxy by means of the use of the tidal mass.

In general, the gravitational potential for a ULIRG is a complex
function of the  mass distribution of the system which evolves with
time but, as a first approximation, it can be considered dominated by
the masses of the main bodies of the interacting galaxies. We will consider
here \emph{two} simple approaches: assuming that the TDG
candidate is suffering the gravitational potential i) of a point mass in the
mass centre and with the total mass of the system ($M^A_{tid}$) and ii) of the
nearest galaxy ($M^B_{tid}$). The 
estimated values are shown in table \ref{props_3}. For $M^A_{tid}$
we used $M =$
\emph{total mass of the system} and $D =$ \emph{distance to the mass
centre}. Masses for the parent galaxies were taken from \citet{col05} and
\citet{gar05}. For the final tidal mass ($M_{tid}$) we have considered the
most conservative approach: the maximum of $M^A_{tid}$ and $M^B_{tid}$. 
It has to
be taken into account that we are measuring the projected distance
(always smaller than the real distance). Therefore, these values have
to be seen as an upper limit to the real tidal mass.

An estimation of the dynamical mass of the candidates is needed to compare
with the tidal masses. For those regions where a reliable measurement of the
velocity dispersion was achieved, this can be done using:

\begin{equation}
M (M_\odot) = cte \times 10^6 R_{\mathrm{eff}}(\mathrm{kpc}) \sigma^2(\mathrm{km~s^{-1}})
\end{equation}

Estimations of the constant range between 1.4 \citep{ben92} and 2.2
\citep{ter81} depending on the method used for its determination and the mass
distribution assumed. Here have adopted  $cte = 2.09$, a value derived from
the Virial theorem, assuming a de Vaucouleurs profile (which describes the
luminosity distribution of spheroidal systems quite well), that luminosity
traces the 
mass of the system \citep{com02} and that the tridimensional velocity
dispersion is related with the observed velocity dispersion as $\langle V^2
\rangle = 3 \sigma^2$ \citep{mcc03}. 
As it can be seen in last column of table 4, most of the candidates
with an estimation of the dynamical mass have a $M_{dyn}/ M_{tid}$ 
ratio which  supports  the idea that most of the regions are
stable against the forces exerted by the parent galaxy (mean $\sim$30). Only
R1 in IRAS 12112+0305 and R2 in IRAS 16007+3743 do not  fulfil the
criterion, but  taking into account the uncertainties involved this is not very
significant even for these two cases.  



A last test could be a comparison between the relative
velocity of a certain region and its escape velocity. 
This criterion cannot be blindly used. Firstly, it is a function of the
distance which will be always larger than the projected one. Secondly, only one
component of the velocity is being measured without having any information
about the movements on the plane of the sky. Finally, as for a certain
configuration it is not possible to determine if the region is closer or
further from the observer than the mass centre, two possibilities for the
relative movements between the region and the system are always possible.
Due to all these uncertainties we cannot give to this comparison the
same importance (weight) as the comparison between the dynamical and tidal
masses. However, we have found useful to include it here for completeness as
used together with other indicators can give us a better idea
about the survival chances of the candidates.

In a simplified way, we consider that the gravitational potential is
created by a point mass in the mass centre and with the 
total mass of the system. 
Relative and escape velocities
can be compared looking at columns 7 and 8 of table \ref{props_3}. As average,
the absolute values of the relative velocities are much smaller than
predicted escape velocities
(189~km~s$^{-1}$ vs 346~km~s$^{-1}$). In order to
take into account projection effects, we will consider that a certain region
will have good survival chances following this test if it satisfy the
condition $\vert v_{rel}\vert - \vert v_{esc} \times \cos(\pi/4) \vert > 0$. According to
this comparison, three out of twelve regions would pass this last test.

\subsection{Comparison with models}

Although the observational characteristics of these
  condensations give some hints that make us to consider them as
  putative TDG progenitors, a comparison with dynamical evolutionary
  models is needed in order to better predict if they will become
  independent objects. 

The formation of TDG inside the tidal tails of interacting galaxies
has been the subject of recent simulations. \citet{wet05} showed the
crucial role of the gas in the formation of these systems, contrary to
the findings of \citet{bar92}. 
\citet{bou06} focused in both their formation and
evolution of this kind of condensations. Typical masses of
the surviving ones are greater than $10^8$~M$_\odot$.
These
simulations end at $\sim$2~Gyr after the beginning of the
encounter. As an average $\sim 3.2$ condensations per galaxy are
formed. From those, only the most massive ($\gsim 10^9$~M$_\odot$) and
born at the tip of the tidal tail ($\sim 0.6 - 1.0$~R$_{tail}$) are
found to survive to the encounter 2~Gyr after the pericenter.
We have performed a comparison of the properties derived for
  our condensations with those created in these simulations.
Seven out of our twelve regions have more than
$10^9$~M$_\odot$. Attending to 
\citet{bou06}, these have more chance to survive to the interaction
process, loosing only a small fraction of their mass.  If the
fate of our regions were similar to the one of these condensations, 
most of them (with the exception of R in IRAS~15250+0305 and R1 in
IRAS~16007+3743) would 
probably fall back to the parent galaxy loosing part of its mass. A
summary of the comparison with these simulations is shown in the sixth
column of table \ref{criteria}.
One question that arise after this comparison is what is the origin of
regions such as R in IRAS~14348$-$1447. Its observational properties
indicate that it is a
condensation of  matter made out of the material of the interaction
which is suffering a burst of star formation and hence, it
should, in principle receive the same treatment as the other 
regions as possible entities able to survive to the merging
process. However, to our knowledge, there is no simulation able to
predict condensations of matter, even transient,  outside the tidal
tails.

How will their evolution be after 2 Gyrs? Unfortunately there are not
evolutionary models covering this phase, for the range of masses of
our regions. However, \citet{kro97} and \citet{met07} simulated the
evolution of condensations with an initial mass of $10^7$~M$_\odot$
(i.e. more than one order of magnitude lower than the present case) up to 10
Gyrs. They show that around $4-6$~Gyr, depending on the model, the
condensations retain  $\sim10$~\% of their initial mass, and only
$\sim1$~\% at $8-10$ Gyr. If these results could be scaled to our more
massive regions, they would end up with masses of about $10^7
-10^8$~M$_\odot$. However, this 
extrapolation may be questionable and a more firm conclusion should
wait for realistic simulations for the appropriate range of mass and time.

\subsection{Summary}

   \begin{table*}
    \caption[]{Summary of the different TDG candidates survival criteria used\label{criteria}}
  \begin{center}
           \begin{tabular}{cccccccccccccc}
             \hline
            \noalign{\smallskip}
Reg.  &
\multicolumn{3}{c}{Observational tests} &
\multicolumn{2}{c}{Comparison with simulations} &
Prob
\\
 &
L(H$\alpha$)     &
$\sigma$ vs. $r$ &
$M_{\mathrm{tid}}$ vs. $M_{\mathrm{cand}}$ &
M$_{lim}$ &
Morph.
\\
            \noalign{\smallskip}
            \hline
            \noalign{\smallskip}
\hline

\multicolumn{7}{c}{IRAS~08572+3915}\\
\hline
R1 & $\surd$ &  \ldots  & \ldots   &  \ldots & $\times$ & Low\\
k7 & $\surd$ &  $\surd$ & $\surd$  &  $\times$ & $\times$ & Medium\\
\hline
\multicolumn{7}{c}{IRAS~12112+0305}\\              
\hline
R1 & $\surd$ &  $\surd$ & $\surd$   &  $\surd$ & \ldots & High-Medium\\
R2 & $\surd$ &  $\surd$ & $\times$  &  $\surd$ & \ldots & Medium\\
kc & $\surd$ &  \ldots  & \ldots    &  \ldots & ? & Low\\
k1 & $\surd$ &  \ldots  & \ldots    &  \ldots & ? & Low\\
k2 & $\surd$ &  \ldots  & \ldots    &  \ldots & ? & Low\\
\hline
\multicolumn{7}{c}{IRAS~14348$-$1447}\\
\hline
R & $\surd$ &  $\surd$ & $\surd$ & $\surd$ & \ldots & High-Medium\\
\hline
\multicolumn{7}{c}{IRAS~15250+3609}\\
\hline
R  & $\surd$ &  $\surd$ & $\surd$ & $\surd$ & $\surd$ & High\\
\hline
\multicolumn{7}{c}{IRAS~16007+3743}  &\\
\hline
R1 & $\surd$ &  $\surd$ & $\surd$  & $\surd$ & $\surd$ & High\\
R2 & $\surd$ &  $\surd$ & $\times$ & $\surd$ & $\times$ & Low\\
R3 & $\surd$ &  $\surd$ & $\surd$  & $\surd$ & $\times$ & High-Medium\\
           \noalign{\smallskip}
            \hline
         \end{tabular}
\end{center}

The second column contains the luminosity 
  criterion. Next column indicates when a certain condensation satisfies the
  relation velocity dispersion-radius of \citet{ter81}. The fourth column
  indicates if the system is stable against the tidal forces exerted by the
  parent galaxy(ies). Next column indicates when the estimated
  dynamical mass is above $10^9$~M$_\odot$. Sixth column show if a
  regions is expected to survive attending to the
  comparison with the simulation of \citet{bou06}. The
  symbol $\surd$ indicates that a certain condensation satisfies a criterion
  while $\times$ indicates that it fails. Symbols with a question mark are
  doubtful. Last column indicates
  the probability that a certain condensation has to become a future TDG based
  on these criteria.
%
   \end{table*}

Due either to projection effects or to observational constrains, 
none of the tests performed above can alone tell us whether a
certain region is going to survive as a future TDG or not. However, a region
that satisfy most of the tests has more chances to survive.

We have summarised the results of these tests in table \ref{criteria}.
We have assigned a  \emph{High} probability of survival to those
regions that satisfy all the utilised criteria (i.e. only R of
IRAS~15250+3609 and R1 in IRAS~16007+3743), \emph{High-Medium} and
\emph{Medium} probability to those that satisfy
four and three of the utilised criteria respectively, and \emph{Low}
probability to those that satisfied only two or one.
The most promising candidate is the one detected in IRAS~15250+3609. The
relative large projected distance to the parent galaxy (6.9~kpc), the
advanced state of the merging process, its position in the $r-\sigma$
plane and the large velocity difference with relation to the parent galaxy
\citep{col05,mon04} make it the best candidate to survive to the merging
process. This region shows the way to be followed
in the hunting of such systems: a systematic search for regions
with recent or ongoing star-formation among a sample
of ULIRGs made out of galaxies in an advanced state of the merging
process seems appealing. The other region that should be mentioned
here is R1 of 
IRAS~16007+3743. Apparently it is at the tip of the tidal tail of a
system that is starting its interaction and depending of its evolution
it could finish in a region similar to the one in IRAS~15250+3609.
There are in addition three candidates with
High-Medium probability (R1 of IRAS~12112+0305, R
of IRAS~14348$-$1447 and R3 in IRAS~16007+3743).
Two of them are new candidates while R
in IRAS~1448$-$1447 had already been proposed as candidate by \citet{mih98}. 
The continuity in the velocity field between them and their 
parent galaxy reduce the chance of survival.

The analysed systems are at higher redshift (and hence worst linear
resolution) than their less luminous relatives previously studied. A
follow-up of these candidates with IFS systems able to provide greater
spatial resolution (i.e. Adaptive Optics assisted systems)
would permit to look for independent velocity gradients in them
similar to the works of \citet{wei02} or \citet{bou04}.

\section{Conclusions}

The present work constitutes the first systematic attempt to try to establish
TDG candidates among ULIRGs.
Extranuclear star-forming regions identified as candidates to TDGs  have been
studied in a sample of low-z Ultraluminous Infrared  Galaxies on the basis of
Integral Field Spectroscopy and high angular resolution HST images. 

As previously found in lower luminosity mergers,  the present work  shows that
the presence of TDG candidates in ULIRGs is common. In  particular we have
identified twelve condensations
  in 5 ULIRGs (from an initial sample of 9), expanding the
luminosity range of interacting galaxies and mergers showing this  phenomenon. 

We have characterised the main physical and kinematic properties of  these
candidates. In particular, their dynamical masses are in the $2 \times10^8 -
1\times10^{10}$~M$_\odot$ range with typical sizes of $ \sim 750$~pc.  We
found that most of these condensations follow the relation between
effective radius and velocity dispersion of lower (globular clusters) and
higher (elliptical) mass systems. 

Starbursts have been identified within each TDG candidate.
Their  \textsc{H\,ii} - like ionisation, relatively high metallicity   ($12 +
\log(\mathrm{O/H}) \sim 8.6$),  and H$\alpha$ equivalent  widths ($34 -
257$~\AA) are characteristic of young bursts of 
star formation ($5-8$~Myr) with stellar masses  between  $2 \times  10^6$ and
$7 \times 10^8$~M$_\odot$.

Using structural, physical and kinematical information, we have  discussed the
likelihood  of survival for these candidates to  internal as well as to the
external forces on the basis of different dynamical tracers. 
Five out of the twelve initial regions present a \emph{High} or
\emph{High-Medium} likelihood of survival as future TDG.
We identified R in IRAS~15250+3609 as our best candidate. It
  satisfies all our tests, occupies in the $r - \sigma$ plane a 
position close to dwarf elliptical galaxies, has relatively high
relative velocity to the parent galaxy and have been found in a system in
an advanced stage of the merging process.

Finally, we have outlined the working lines that need to be followed
in the confirmation of these candidates as well as when looking for
new ones. That is: IFS observation with high spatial resolution as
those provided by Adaptive Optics assisted IFS systems and the
systematic search of external regions of star formation in ULIRGs in
an advanced state of the merging process.


\begin{acknowledgements}

The authors wish to thank P.~Weilbacher and L.~M.~Cair\'os as well as
the referee, P.-A.~Duc for useful comments 
and suggestions which helped to improve the paper.
AMI acknowledges support from the Euro3D Research Training Network,
funded by the EC (HPRN-CT-2002-00305). Financial support was provided
by the Spanish Ministry for Education and Science through grant
AYA2002-01055. Work based on observations with the William Herschel Telescope
operated on the island of La Palma by the ING in the Spanish Observatorio del
Roque de los Muchachos of the Instituto de Astrof\'{\i}sica de Canarias.

\end{acknowledgements}

\end{document}